\documentclass[aps,prl,reprint,superscriptaddress]{revtex4-2}

\usepackage[T1]{fontenc}
\usepackage[utf8]{inputenc}
\usepackage{url}
\usepackage{amsmath}
\usepackage{mathtools}
\usepackage{bbold}
\usepackage{mathrsfs}

\usepackage{graphicx}
\usepackage{epstopdf} 
\usepackage{wrapfig}
\usepackage{braket}
\usepackage{xcolor}
\usepackage[font=small,skip=0.1pt,,justification=raggedright]{caption}

\begin{document}
 
\author{Giovanni Pecci}
\affiliation{Univ. Grenoble Alpes, CNRS, LPMMC, 38000 Grenoble, France}

\author{Piero Naldesi}
\affiliation{Univ.  Grenoble Alpes, CNRS, LPMMC, 38000 Grenoble, France}

\author{Luigi Amico}

\affiliation{Quantum Research Centre, Technology Innovation Institute, Abu Dhabi, UAE}

\affiliation{CNR-IMM $\&$ INFN-Sezione di Catania, Via S. Sofia 64, 95127 Catania, Italy}
\affiliation{Centre for Quantum Technologies, National University of Singapore, 3 Science Drive 2, Singapore 117543, Singapore}
\affiliation{LANEF Chaire d’excellence, Univ. Grenoble-Alpes $\&$ CNRS, F-38000 Grenoble, France}
\author{Anna Minguzzi}
\affiliation{Univ. Grenoble Alpes, CNRS, LPMMC, 38000 Grenoble, France}

\title{Probing the BCS-BEC crossover with persistent currents}


\begin{abstract} 
We study the persistent currents of an attractive Fermi gas confined in a tightly-confining ring trap and subjected to an artificial gauge field all through the BCS-BEC crossover. At weak attractions, on the BCS side, fermions display a parity effect in the persistent currents, ie their response to the gauge field is paramagnetic or diamagnetic depending on the number of pairs on the ring. At resonance and on the BEC side of the crossover we find a doubling of the periodicity of the ground-state energy  as a function of the artificial gauge field and disappearance of the parity effect,  indicating that persistent currents can be used to infer the formation of tightly-bound bosonic pairs. Our predictions can be accessed in ultracold atoms experiments through noise interferograms.
\end{abstract}

\maketitle

\paragraph*{Introduction}
A gas of weakly attractive spin one half fermions can form bound pairs with opposite spin and condense into the Bardeen-Cooper-Schrieffer (BCS) paired regime. On the other hand, particles with integer spin can display Bose-Einstein condensation (BEC). 
Despite BCS pairing and BEC are two distinct physical phenomena, they have been intensively studied as two different regimes that may occur in the same system. In the BCS regime, the correlation length of the bound pairs is large compared with the typical interparticle distance; in the BEC regime, instead, the pairs are tightly bound in the real space and the pair  correlation length is much smaller than the distance between the particles. The evolution between the two regimes is called BCS-BEC crossover\cite{strinati2018bcs}. It plays an important role in different contexts, ranging from nuclear\cite{baldo1995deuteron,yang2000recent} to condensed matter physics\cite{randeria1989bound,micnas1990superconductivity}. With the advent of ultracold atoms quantum technology, the BCS-BEC crossover has been studied in the laboratory with unprecedented control of the physical conditions, making it possible to test important aspects of the different theories developed so far \cite{randeria2012bcs,inguscio2008ultra,Valtolina_2015}.

\begin{figure*}\centering \includegraphics[width=1\textwidth]{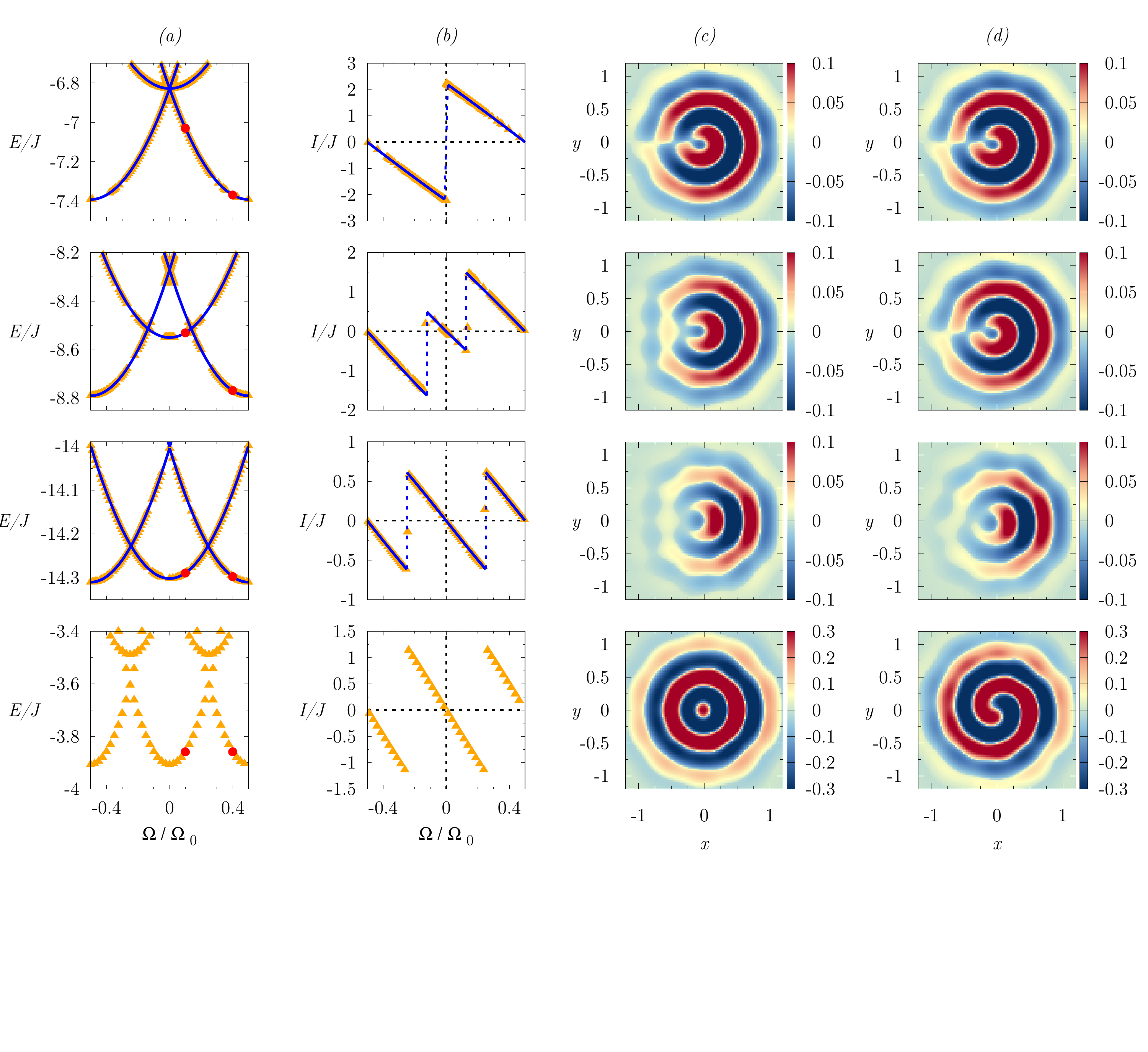} 
\vspace{-3cm}
\caption{(Color online) Column \textit{(a)}: energies vs $\Omega$ for $N=4$, $N_s = 8$  from Bethe Ansatz (blue line) and DMRG (yellow triangles). In each column, the first three panels from the top correspond to $U/J = 0, -2, -6$ respectively. The last line describes the BEC regime for $U_B/J_B =1$ -- here only DMRG data are available since the model is not integrable. Column \textit{ (b)}: persistent current as a function of the flux for the same set of parameters,  obtained as the derivative of the ground-state energy in column \textit{ (a)}. Columns \textit{(c)} and \textit{(d)}: noise correlator for $N=4$, $N_s = 10$ for $\Omega/\Omega_0 = 0.1$ and $\Omega/\Omega_0 = 0.4$ respectively, indicated by red circles in column \textit{(a)}.  The correlators are all evaluated in $r' = (R, 0)$ and $t = 0.6 \ m R^2/\hbar N_s$. A circulating state is characterised by a spiral-like  correlator, not symmetric by inversion with respect to the $y=0$-axis.  } 
\label{N4} 
\end{figure*}

\begin{figure*}
\centering
\includegraphics[width=1\textwidth]{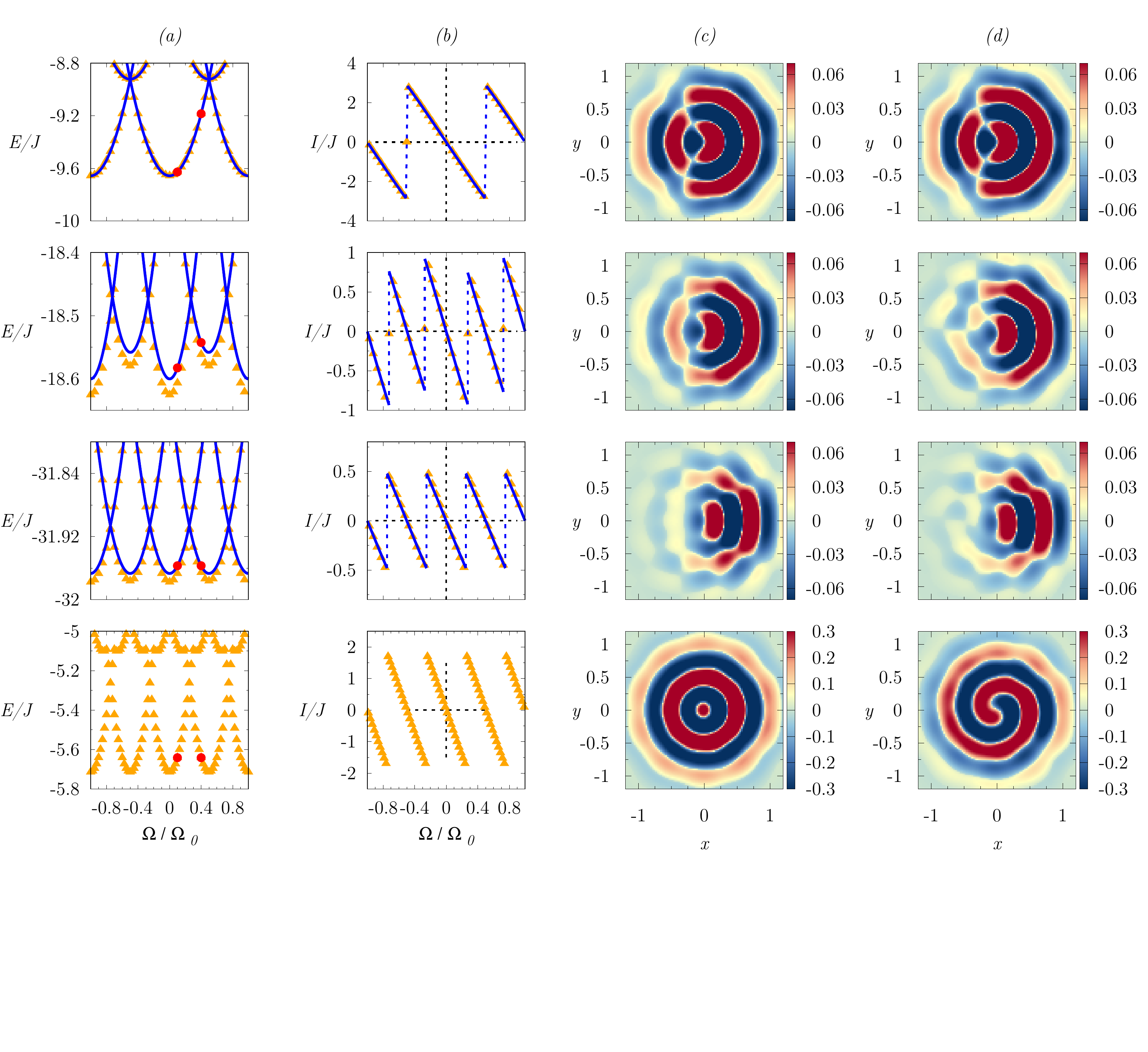}
\vspace{-3cm}
\caption{(Color online) Energies, persistent current and noise correlators as in Figure \ref{N4} here studied for $N=6$, and for  $U/J = 0, -5, -10$ (from top to bottom)  \cite{note}. The last line refers to the bosonic case for $U_B/J_B=1$. All the other parameters are the same as in Fig.~\ref{N4}. 
} 
\label{N6}
\end{figure*}

Recently, important progress has been achieved in the field, allowing coherent manipulation of atoms in trapping potentials with wide ranges of intensities and shapes, in an unprecedented precise manner\textcolor{black}{\cite{kwon2020strongly,rubinsztein2016roadmap,brantut2012conduction}}.
Atomtronics exploits such remarkable progress to realize matter-wave circuits of ultracold atoms manipulated in magnetic or lasers-generated guides\cite{amico2020roadmap,Amico_NJP,dumke2016roadmap}. In particular, harnessing current states in an explicit way, atomtronics effectively widens the scope of quantum simulators and emulators to probe quantum phases of matter. Specifically, in the spirit of solid-state physics I-V (current-voltage) characteristics, the different many-body quantum regimes are characterized in terms of the current flowing through the cold atomic system. Here, we take the latter view to study the BCS-BEC crossover: we show how the persistent current in attracting fermionic systems confined in ring-shape potential and pierced by an artificial  gauge field provides a novel way to tell apart the BCS and BEC regimes. 
As in \cite{fuchs2004exactly}, we describe the system on the BCS side using a model of fermions with attractive contact interactions. At resonance, reached  in the infinitely strongly attracting limit, it corresponds to  a Tonks-Girardeau gas of hardcore bosonic pairs \cite{iida2005exact}. The BEC side of the crossover is described by a bosonic model with contact interactions for the pairs. By applying exact Bethe Ansatz methods corroborated by Density Matrix Renormalization Group (DMRG) simulations, we access to all regime of
interactions ranging from weak to strong attractions. 
Our main results are summarized in Fig. \ref{N4} and \ref{N6}. 
We rely on a theorem due to Leggett\cite{nanoelectronics1991dk} to demonstrate that the BCS to BEC crossover is marked by clear features of the periodicity of the persistent currents.
Accordingly, assuming that the total number of particles is $N=2n$, the persistent current of a gas of interacting spin half fermions is predicted to be parity dependent: for even number $n$ of pairs of fermions, the system behaves as a paramagnet with a non-vanishing persistent current at zero effective magnetic field; for odd $n$ instead, the system behaves as a diamagnet (vanishing persistent current at zero field). 
For our specific problem, we find that while the persistent current displays clear parity dependence in the BCS regime, the latter is washed out for strongly attracting pairs, indicating that at resonance and in the BEC regime, fermionic pairs behave as point-like bosons, which are predicted not to show parity effects \cite{schilling2016number,manninen2012quantum}.
Finally, inspired by a procedure developed for bosonic condensates \cite{corman2014quench,wright2013driving,eckel2014interferometric,mathew2015self}, we  propose a protocol to evidence the parity effect in the persistent current by noise correlations, based on  the self-heterodyne detection of the phase of the many-body wavefunction.

\paragraph*{The model}
We consider a gas of degenerate fermions confined in a lattice ring  
of radius $R$ and
pierced by an effective gauge field. Such system is  a paradigmatic example of atomtronic circuit \cite{Amico_NJP,dumke2016roadmap}. 
The artificial magnetic field can be applied in several ways, for example by stirring the condensate, by phase imprinting or two-photon Raman transitions \cite{dalibard2011colloquium}.
To address the whole BCS-BEC crossover, we exploit the possibility to tune the interaction strength in ultracold atoms, eg across a confinement-induced resonance \cite{PhysRevLett.81.938,valiente2011quasi}. 
To describe the BCS side of the crossover, we use the one-dimensional attractive Fermi-Hubbard Hamiltonian, which, for an even number $N$ of particles on a lattice of $N_s$ sites, reads
\begin{equation}
\small{\hat{\mathcal{H}}_{FH} = -J \sum_{j=1}^{N_s}\sum_{\sigma =\uparrow,\downarrow} \Bigl(e^{i\frac{2\pi }{Ns}\frac{\Omega}{\Omega_0}}c^\dagger_{j, \sigma}c_{j+1, \sigma} + \textit{H.c.} \Bigr) + U \sum_{j=1}^{N_s}n_{j,\uparrow} n_{j,\downarrow} }
\label{Hubbard}
\end{equation}
where $U<0$, $J$ is the tunnel amplitude, $\Omega$ is the rotation frequency induced by the artificial gauge field and $\Omega_0=\hbar/mR^2$. The Hamiltonian (\ref{Hubbard}) is solvable by Bethe Ansatz\cite{lieb1968absence,ShastryFlux} with many-body eigenvalues in the form  
$ E = -2 \sum_{j} \cos k_j \label{energy} \;$. The quantities $k_j$ are the so-called charge rapidities: we set them as
$k_{j,\pm}=p_j \pm i v_j$. In thermodynamic limit they correspond to two-dimensional \textit{k}-strings in which all the charge rapidities are in the form $\sin(k_{j,\pm})=\Lambda_j\pm i U/4J$. A non-vanishing imaginary part of the rapidities indicates the formation of bound states. Excitations on top of the ground states are obtained either by pair breaking (gapped spin excitations), by pair re-arrangements (gapless charge excitations), or by combinations of thereof \cite{andrei1995integrable}. We note that solutions deviating from the \textit{k}-string form  occur if $U N_s/J$ is not sufficiently large. This may happen on a lattice ring at very weak interactions. We provide a solution to this model at any interaction strength and in the total spin zero sector. 

For strongly attractive interactions, the Bethe Ansatz solution indicates the formation of a Tonks-Girardeau gas of hard-core bosons. Hence, in the spirit of \cite{fuchs2004exactly}, we describe the BEC side of the crossover by a bosonic Hubbard model, where bosons correspond to the fermionic pairs,
\begin{equation}
\small{\hat{\mathcal{H}}_{BH} = -J_B \sum_{j=1}^{N_s} \Bigl(e^{i\frac{2\pi }{Ns}\frac{\Omega}{\Omega_0}}b^\dagger_{j, \sigma}b_{j+1, \sigma} + \textit{H.c.} \Bigr) + \frac{U_B}{2} \sum_{j=1}^{N_s}n_{j} (n_{j}-1) }.
\label{bHubbard}
\end{equation}
Here $J_B$ is the tunnel amplitude for pairs and $U_B$ is the pair-pair  interaction strength \cite{SupplMaterial}. We remark that at difference from the Fermi-Hubbard model, the bosonic one is not integrable \cite{choy1982failure,choy1980some}: there is  no exact solution for this model.
To address the whole BCS-BEC crossover, in our work we combine  Bethe Ansatz calculations and  DMRG simulations to calculate ground and excited state energy branches as a function of $\Omega$. DMRG is also used to calculate noise correlators. 

\paragraph*{Parity effects of the persistent current.}
We obtain the persistent current from the ground-state energy branch according to $I=- \Omega_0\partial E_{GS}/\partial \Omega$.
For zero and weak interactions, the persistent current is a periodic function of the flux with period $\Omega_0$. Its behaviour markedly depends on the parity of the number of pairs $N/2$ (see Figs.\ref{N4} and \ref{N6}): for {\it even} $N/2$, the ground state energy displays a global maximum for $\Omega = 0$ (paramagnetic behaviour); for {\it odd} $N/2$, instead, the ground state energy has a global minimum for $\Omega = 0$ (diamagnetic behaviour). Such effect emerges by comparing the first two columns of Fig.\ref{N4} and \ref{N6}. Remarkably, {\it the parity dependence of the persistent current  disappears for strong enough interactions}:  upon increasing  interactions the ground-state energy displays a superlattice structure  as the energy of the excited states decreases, leading to the doubling of its periodicity and the suppression of the parity effects in the persistent current at resonance. 

On the BEC side the periodicity of the ground state energy is $\Omega_0/2$, corresponding to the quantum of flux of a pair, for any $U_B$. Our results  elucidate the mechanism of doubling of the periodicity originally predicted by Byers and Yang for superconductors \cite{ByersYang}.
The decrease of the energy of the excited states is  also at the origin of the change of sign in the curvature of the ground state energy at zero flux observed in  the  disordered Fermi-Hubbard model for even $N/2$ \cite{waintal2008persistent}: disorder smooths the cusps at weak interactions yielding a negative curvature, while a positive curvature is found  at strong attractions as in the clean case.

\paragraph*{Readout: interferograms and noise correlations.}
We next  suggest  a protocol to probe the BCS-BEC crossover in cold atoms set up. In analogy to the approach carried out for bosonic systems, we propose an interferometric detection of the current by studying the interference pattern arising from the co-expansion of the gas on the ring and a degenerate gas placed at the center $C$ of the ring\cite{SupplMaterial}. We model the gas in the center as a single site, and assume that no hopping between the ring and the central site can occur.
We  then study density-density correlations at equal time, which on the BCS side reads $\sum_{\rho, \sigma = \uparrow, \downarrow}\langle n_\rho(r,t)n_\sigma(r',t)\rangle$, where $n_\sigma(r,t) = \Psi_\sigma^\dagger (r,t) \Psi_\sigma (r,t)$, is  the  density operator for the spin component $\sigma$, $\Psi_\sigma(r,t)$ being the fermionic field operator.  On  the BEC side the first non trivial correlator is the one associated to density of pairs $n_B$  \cite{haug2018readout}, ie $\langle n_B(r,t)n_B(r',t)\rangle$. 

In order to enhance the visibility of the correlator, arising from the low density of the system, we generalize the method  devised in \cite{haug2018readout}:  we find that the only terms producing a non trivial interference pattern are the ones describing the correlations between the expanding ring and center:
\begin{align}
&\tilde{G}(r,r',t)= -\tilde{G}_0(r,r',t) +\sum_{\rho, \sigma = \uparrow, \downarrow} \left[
\langle n_\rho(r,t)n_\sigma(r',t)\rangle \notag \right. 
\\ & \left. -  \langle n_\rho(r,t)\rangle_{\text{ring}} \langle n_\sigma(r',t)\rangle_{\text{C}} 
-\langle n_\rho(r,t)\rangle_{\text{C}} \langle n_\sigma(r',t)\rangle_{\text{ring}} 
 \notag \right. \\ 
& \left. -\langle n_\rho(r,t) n_\sigma(r',t)\rangle_{\text{ring}} -  \langle n_\rho(r,t) n_\sigma(r',t)\rangle_{\text{C}}
\right]
\label{corr1}
\end{align}
where  $\tilde{G}_0(r,r',t)=w^*_C(r,t) w_C(r',t) \sum_{j=1}^{N_s} w_j(r,t)w^*_j(r',t)$ has also  been subtracted, with $w_j(r,t)$ the expanding Wannier function on the site $j$ at time $t$ \cite{SupplMaterial} and we have set $\Psi_\sigma(r)=w_C(r) c_{C,\sigma}+\sum_{j=1}^{N_s} w_j(r) c_{j,\sigma}$. 
The $\tilde{G}_0$ term corresponds to the  one-body density matrix of a non-interacting Fermi gas for a completely filled lattice, ie when it forms a band insulator.
On the BEC side, a definition analogous to Eq.(\ref{corr1}) involving the bosonic density is used.
We remark that the above scheme leads to a priori non-Hermitian $\tilde{G}(r,r',t)$ \cite{SupplMaterial}, hence in the following, we focus only on square root of the real part of $\tilde{G}(r,r',t)$, ie $G(r,r',t) = \text{sgn}\Bigl(\text{Re}[\tilde{G}(r,r',t)]\Bigr)\sqrt{\text{Re}[\tilde{G}(r,r',t)]}$.

We study the interference pattern  for systems with even and odd number of pairs in columns (\textit{c}) and (\textit{d}) of Figs.~\ref{N4} and \ref{N6}. 
In the small $\Omega$ case, at weak interactions  we see a spiral  image with a dislocation indicating non-zero current for even $N/2$, while there is no current for odd $N/2$ as the figure is symmetric by reflection along the $y=0$ axis. The shape of the interferograms is  due to the different contributions of the single-particle orbitals constituting the Fermi sphere \cite{phase_prep}. In contrast, at strong interactions, the
images for $N/2$ even or odd  are both symmetric with respect to the $y=0$ axis, indicating that  the  current vanishes regardless of the parity of the number of pairs.
We also study the correlation in the 
interference pattern for $\Omega/\Omega_0$ slightly below the degeneracy point $\Omega/\Omega_0=1/2$. Comparing the columns (\textit{c}) and (\textit{d}) of both Fig.\ref{N4} and \ref{N6}, we see that at zero interactions the current is the same in the two cases, consistently with the fact that the two values of flux are on the same period of the ground state energy. 
At stronger interactions, close to resonance,  the images are more blurred by the reduced phase coherence. Nevertheless, we see that the correlation functions of column (d) display non-mirror symmetric, spiral-like features, indicating the presence of a current state and the doubling of the periodicity.

\paragraph*{Conclusions and outlook}
We studied the persistent current of a Fermi gas  confined in a mesoscopic ring-shaped lattice subjected to an  artificial magnetic field all through the BCS-BEC crossover. We described the system through attractive Fermi  and repulsive Bose Hubbard models, using both the exact solution by Bethe Ansatz and DMRG.

We demonstrated that the persistent current displays distinctive features in the various interaction regimes.
At weak interactions (BCS regime), the persistent current is a periodic function of the single-particle flux quantum, displaying some modulations due to the superlattice structure of the ground-state energy. Such phenomenon indicates the onset of pairing correlations building up between up and down spins. The BCS  regime is  characterized by parity-dependent persistent currents: while for odd number of pairs the system has a diamagnetic response, for even number of pairs the system has paramagnetic response. Remarkably, the parity effect is washed out at resonance  and in the BEC regime. In these regimes, the persistent current is periodic with a two-particle flux quantum, providing a clear signature of 
the formation of bound pairs. To experimentally monitor the features of the persistent current described above, we let the gas co-expand with a reference gas placed in the center and we study the noise correlations in the interference pattern. Quite remarkably, such interferometric analysis works also in the BCS regime, where the phase coherence ensured  by the fermionic pairs is lower than in the BEC one.

Our work provides a clear evidence  that the response of mesoscopic size system to an artificial gauge field is a relevant tool to study the BCS-BEC crossover.
Our approach is fully general and readily applicable to other models, eg the  boson-fermion one \cite{tianhaoaleiner} as well as to other systems, such as high $T_c$ superconductors \cite{chen2006applying, uemura1997bose}.

\paragraph*{Acknowledgements}
We thank T. Haug and C. Salomon for discussions. The Grenoble LANEF framework ANR-10-LABX-51-01 are acknowledged for their support with mutualized infrastructure.

\bibliography{BCS_BEC_crossoverARXIV.bib}

\begin{thebibliography}{49}
\bibitem [{not()}]{note}%
  \BibitemOpen
  \href@noop {} {}\bibinfo {note} {The slight discrepancy between DMRG and
  Bethe Ansatz results for $U/J=-5$ is due to the string hypothesis made in the
  latter.}\BibitemShut {Stop}%
\end{thebibliography}

\appendix

\section{SUPPLEMENTAL MATERIAL}

\maketitle

\subsection{Solution of Bethe equations}

The Bethe equations for the Fermi Hubbard model read\cite{lieb1968absence}:
\begin{align}
&\small{e^{i N_s k_j}= e^{2\pi i\Omega/N_s}} \notag \prod_{l=1}^M \frac{-\Lambda_l+\sin(k_j)+\frac{i U}{4}}{-\Lambda_l+ \sin(k_j)-\frac{i U}{4}}    \\
&\small{\prod_{j=1}^N\frac{-\Lambda_\alpha+ \sin(k_j)+\frac{iU}{4}}{-\Lambda_\alpha + \sin(k_j)-\frac{i U}{4}} = \prod_{\substack{\beta=1 \\ k \neq m}}^M\frac{-\Lambda_\alpha+\Lambda_\beta+\frac{i U}{2}}{-\Lambda_\alpha+ \Lambda_\beta-\frac{i U}{2}}} 
\label{BetheEq}
\end{align}
where $M$ is the number of spin down particles. The variables  $k_j \ j=1... N$ are  the  charge rapidities and $\Lambda_\alpha, \ \alpha=1... M$ as the spin rapidities.  In order the Bethe equations to have  well defined solution, the rapidities must be distinct from each other. The charge rapidities fix the center of mass momentum and the energy eigenvalues, respectively 
$P = \sum_{j=1}^{N} k_j$ and $E = -2\sum_{j=1}^N \cos(k_j)$.

In order to solve Eq.(\ref{BetheEq}) we implement an iterative method: we compute the solution at fixed interaction $U_0$ and then use this value as initial condition for a root finding algorithm to determine the solution at increasing attractions. We choose as initial step the solution for $U= 0$, that can be easily obtained analitically. Finally, to further benchmark our results, we compare the Bethe Ansatz solution and the DMRG calculation at different interaction regimes.

\begin{figure}
\centering
\includegraphics[width=0.5\textwidth]{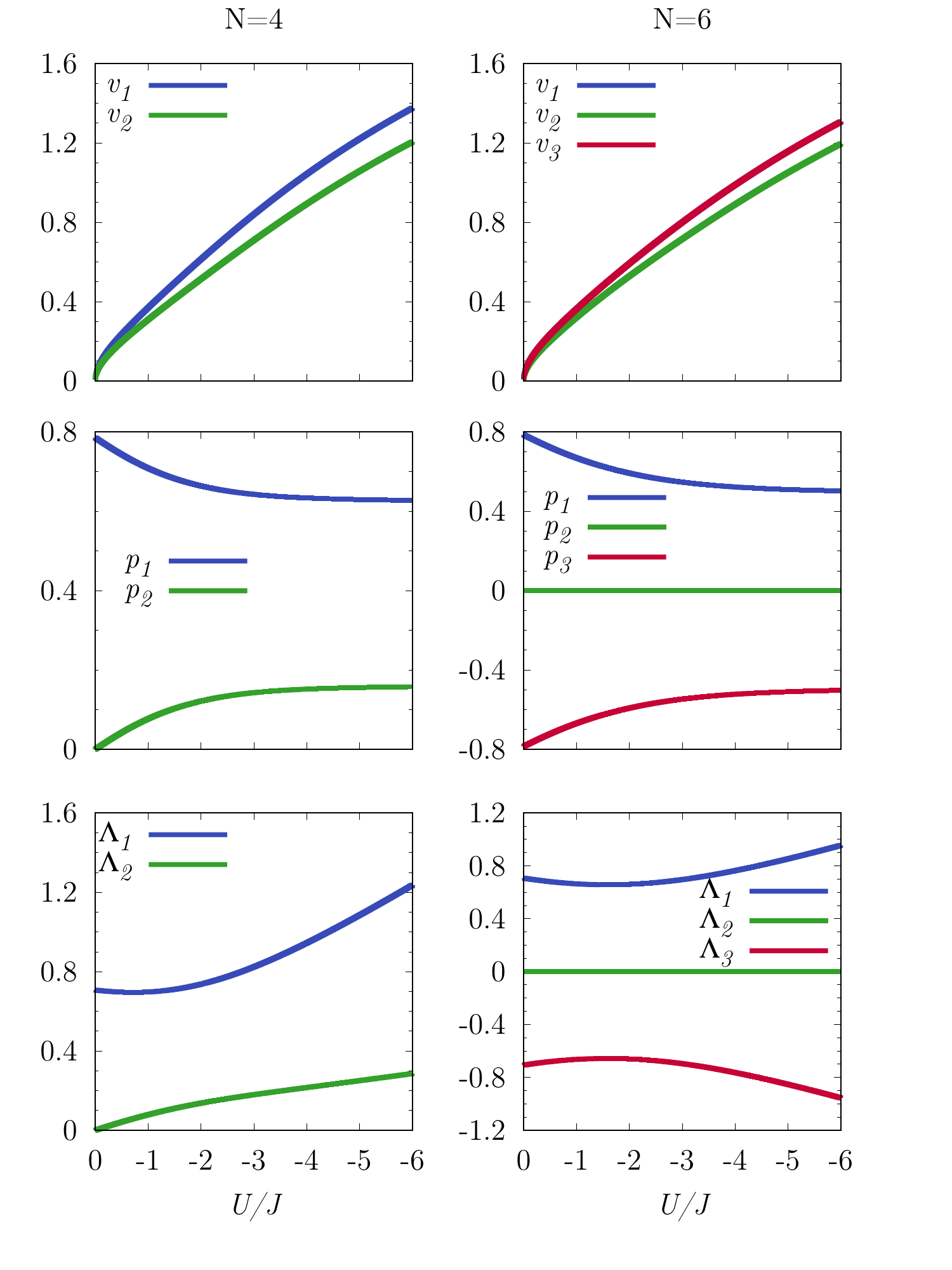}
\caption{\small{Solutions of the Bethe equations as a function of interaction for $N_s = 8$, $N=4$ and $M=2$ (above) and $N_s = 8$, $N=6$, $M=3$ (below). The $N=4$ column corresponds to the branch centered in $\Omega/\Omega_0 = 0.5$ in column (\textit{a}) of Fig.$1$ in the main text, while the $N=6$ panels corresponds to the one centered in  $\Omega/\Omega_0 = 0$ in the same column of Fig.2.}}
\label{Fig.rapidities}
\end{figure}

\begin{figure}
\centering
\includegraphics[width=0.5\textwidth]{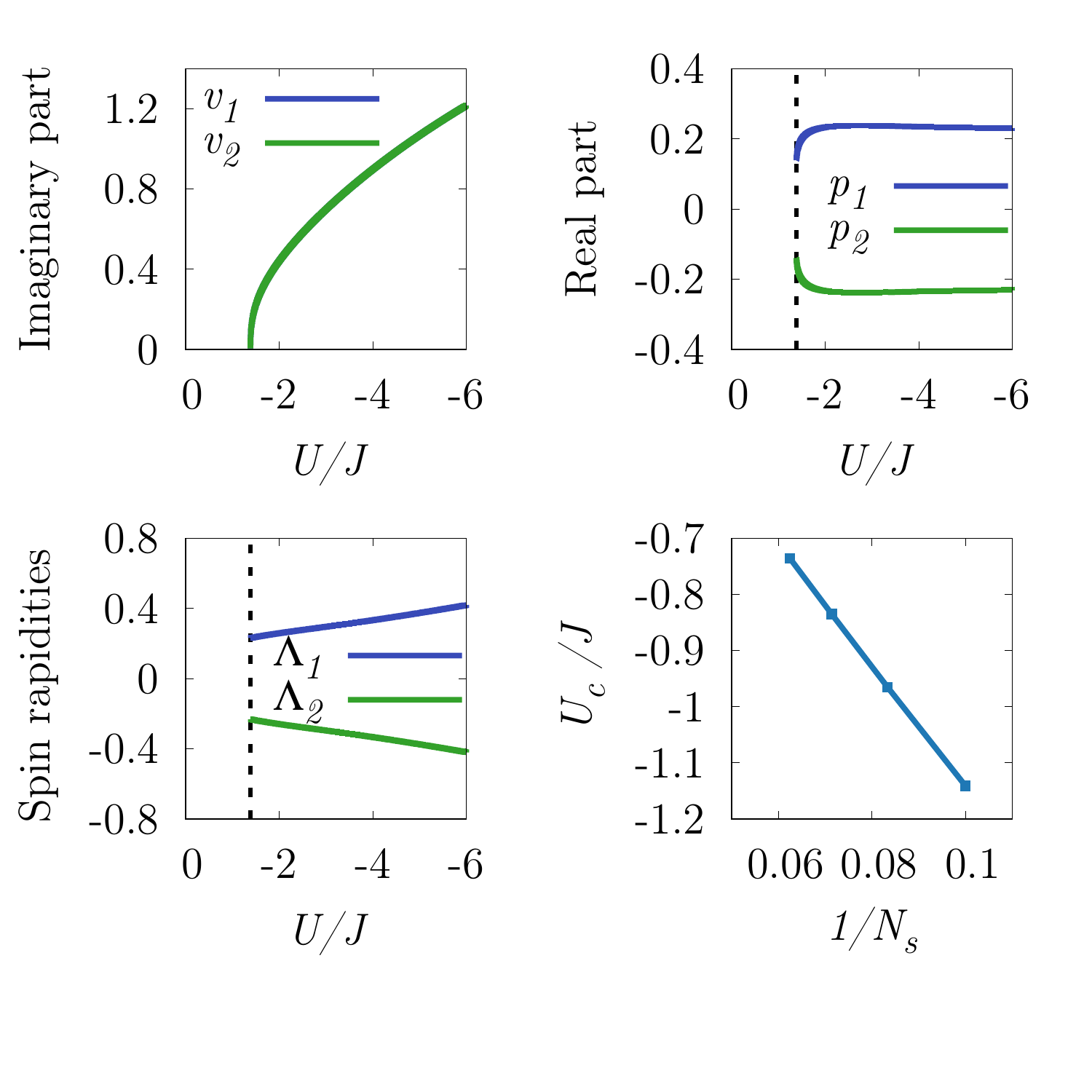}
\caption{(Color online) Solution of the Bethe Ansatz equations: charge and spin rapidities (dimensionless) as a function of interaction $U$ (in units of $J$)  for $N=4$ and $N_s = 8$,  and scaling of $U_c/J$ with number of lattice sites $N_s$. For $U<U_c$ the imaginary part of the rapidities, thus the binding energy, vanishes. This solution corresponds to the energy branch yielding the doubling of the periodicity at large $|U|$. }
\label{cri}
\end{figure}

In Fig.\ref{Fig.rapidities} we show an example  of solutions of Bethe equations for the Hubbard model as a function of dimensionless coupling constant $U/J$. For $U=0$ and $\Omega=0$ the rapidities corresponding to the ground state of the system  have  a center of mass momentum   $P=4 \pi / N_s$  for $N=4$ and  $P=0$ for $N=6$. Other branches of the excitation spectrum are obtained by choosing different values of the center of mass momentum. 

Another type of solutions is shown in Fig. \ref{cri}.  These correspond to the energy branch centered in $\Omega/\Omega_0 = 0$ in Fig $1$, ie the branch responsible for the doubling of periodicity. Remarkably, in this case, the charge rapidities have an imaginary part only for finite interactions $|U| > |U_c|$, where $U_c$ depends on the density of the gas, showing that this branch is not connected with the non-interacting solution. We also find that  $U_c$ decreases at increasing system  size.

\subsection{Mapping from Fermi to Bose Hubbard}
In this section we discuss the relations between the coupling parameters in the Fermi and in the Bose Hubbard Hamiltonians. These are the lattice regularization of the Gaudin-Yang and the Lieb-Liniger models respectively.
The first one describes a one dimensional gas of $N_F$ fermions with contact interactions confined on a homogeneous ring of radius $R$: the Hamiltonian reads\cite{gaudin1967systeme,yang1967}:
\begin{equation}
\hat{\mathcal{H}}_{GY} = -\frac{\hbar^2}{2m_F}\sum_{i=1}^{N_F} \frac{\partial^2}{\partial x_i^2} + g_F \sum_{i \neq j} \delta(x_i - x_j)
\end{equation}

\begin{figure}
\centering
\includegraphics[width=0.45\textwidth]{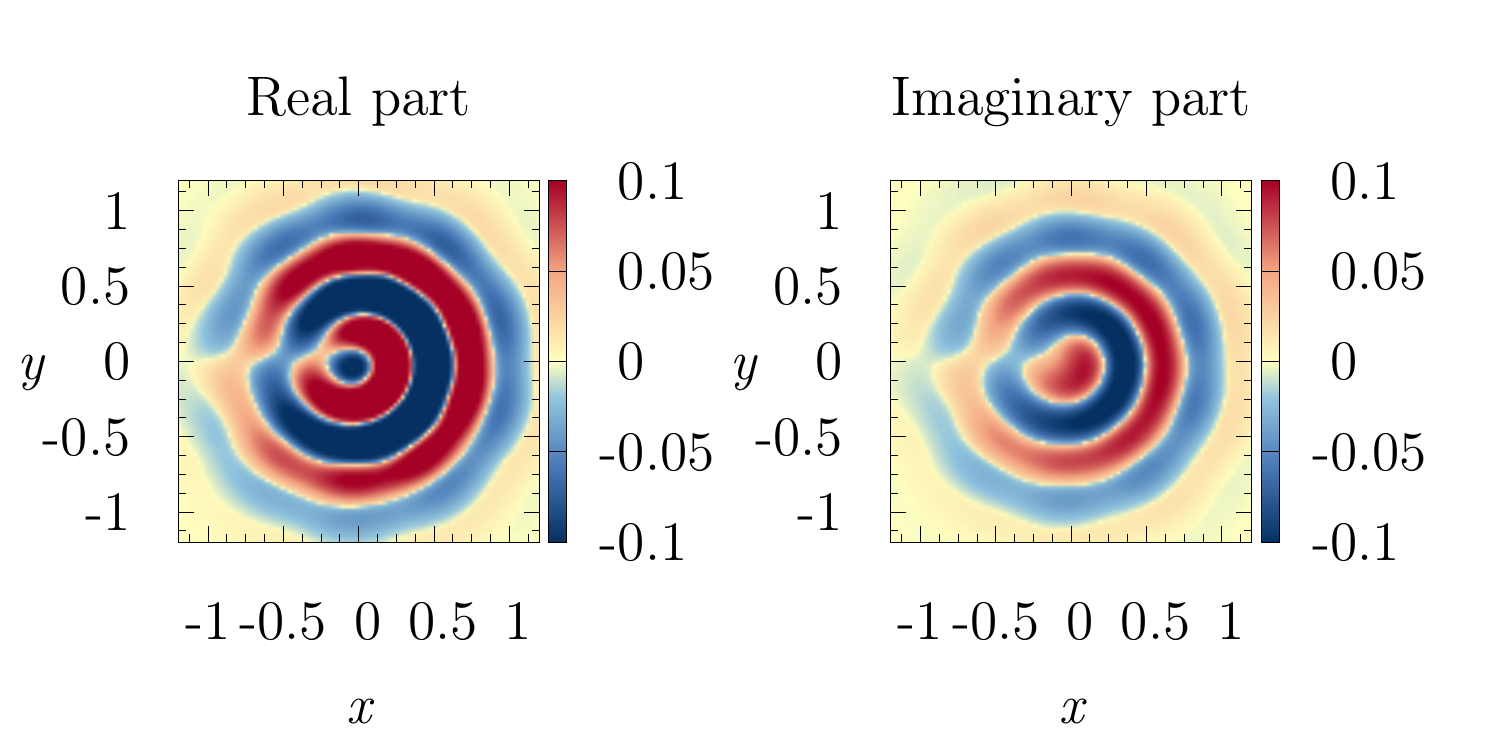}
\caption{Comparison between real and imaginary part of the density-density correlator, calculated for $N=4$ , $N_s=10$, $U=0$, $\Omega/\Omega_0 = 0.4$ and $t = 0.6 \ m R^2/\hbar N_s$. The order of magnitude of the two functions is the same and they display similar topological features. }
\label{RealIm}
\end{figure}

This model is solvable using Bethe Ansatz: the energy eigenvalues can be computed at any interaction strength. Introducing the density of the gas as $n_F = N_F/(2\pi R)$ one can express the coupling using a dimensionless quantity $\gamma_F = \frac{m_F g_F}{\hbar^2 n_F}$. 
The Lieb-Liniger model is an integrable model describing the same setup, but for a gas of $N_B$ bosons. The Hamiltonian can be written in a similar way\cite{lieb1963exact}:
\begin{equation}
\hat{\mathcal{H}}_{LL} = -\frac{\hbar^2}{2m_B}\sum_{i=1}^{N_B} \frac{\partial^2}{\partial x_i^2} + g_B \sum_{i \neq j} \delta(x_i - x_j)
\end{equation}
The coupling can be equivalently expressed in dimensionless unit using $\gamma_B = \frac{m_B g_B}{\hbar^2 n_B}$, where $n_B = N_B/(2\pi R)$ is the density of the Bose gas. It can be shown \cite{iida2005exact, gaudin2014bethe} that in thermodynamic limit the ground state energies of the two model can be mapped one into the other when $\gamma_F \ll - 1$ and $\gamma_B \gg 1$. Such mapping also implies a rescaling of the densities and of the masses of the particles such that $n_B = n_F/2$ and $m_B = 2m_F$, implying that $\gamma_B = 4 |\gamma_F|$. 

When we discretize the Gaudin-Yang model into the Fermi Hubbard, the dimensionless coupling parameter $U/J$ is related to the one of the continuous theory by $\gamma_{F} = \frac{N_s}{N_F} \frac{U}{J}$. An analogous relation holds for the Lieb-Liniger and the Bose Hubbard models \cite{polo2020exact}. From the above considerations, we exstimate the scaling of the dimensionless parameter in the mapping between the fermionic and the bosonic model proposed in the main text. Taking into account the fact that in the bosonic side of the mapping we consider $N/2$ dimers of fermions with double mass $2m$, we obtain $U_B/J_B = 2 |U|/J$.

\subsection{Details on the interference for expanding ring and disk}
In order to describe the expansion after releasing of the confinement we use a Gaussian approximation for the Wannier functions \cite{slater1952soluble, chiofalo2000collective} and the exact solution for the expansion of a quantum particle by releasing it from a harmonic potential \cite{minguzzi2005exact}. This yields \cite{haug2018readout}
\begin{equation}
w_j (r,t) = \frac{1}{\sqrt{\pi}\sigma } \frac{1}{(1 + i \omega_0t)} \exp\Bigl\{-\frac{(r - r_j)^2}{2\sigma^2(1+ i \omega_0 t)}\Bigr\}
\label{wannier}
\end{equation}
where $\sigma=\sqrt{\hbar/m \omega_0}$ and $r_j$ are the initial width and the center of the $j$-th Wannier function respectively, $\omega_0$ being the frequency of the bottom of each lattice well in the harmonic approximation. 

We notice that this approximate expression does not satisfy exactly the completeness relation for the fermionic field at all times. Hence, the noise correlator is in the general case complex, however it becomes real at long times. \cite{altman2004probing}.  

In the main text, we have chosen to show the real part of the correlator  $\tilde{G}(r,r',t)$ at intermediate times. A comparison between imaginary and real part of the correlator is presented in Fig.\ref{RealIm}: we see that essentially both real and imaginary parts  carry the same physical information, this justifying our choice. 

\end{document}